\documentclass[12pt,preprint]{aastex}
\usepackage{graphics}
\usepackage{aastexug}  

\def\Msun{M$_\odot$}
\def\msun{M$_\odot$}

\def\Mprogl{M$_{prog1}$}
\def\Mprogu{M$_{prog2}$}
\def\Mtrans{M$_{trans}$}

\def\Mv{M$_{\rm v}$} 
\def\Teff{T$_{\rm eff}$} 
\def\teff{T$_{\rm eff}$} 
\def\0BMV{$(B-V)_{\rm 0}$} 
 
\def\Z{Z} 
\def\Y{Y}

\def\simgt{\lower.5ex\hbox{$\; \buildrel > \over \sim \;$}} 
\def\simlt{\lower.5ex\hbox{$\; \buildrel < \over \sim \;$}}
\begin{document} 
%
\title{The early evolution of Globular Clusters: the case of NGC 2808}
\author{Francesca D'Antona\altaffilmark{1}  and Vittoria Caloi \altaffilmark{2} } 
%
%
\affil{\altaffilmark{1}INAF - Osservatorio Astronomico  di Roma, via Frascati
33, 00127 Roma, Italy; dantona@mporzio.astro.it}
\affil{\altaffilmark{2}
Istituto di Astrofisica Spaziale e Fisica Cosmica, C.N.R., Via 
Fosso del Cavaliere, 
I-00133 Roma, Italy; caloi@rm.iasf.cnr.it }

 
\begin{abstract}

Enhancement and spread of helium among globular cluster stars have been recently 
suggested as a way to explain the horizontal branch blue tails, in those 
clusters which show a primordial spread in the abundances of CNO and other 
elements involved in advanced CNO burning. Helium enhancement is unavoidable,  if 
the matter responsible for the abundance spreads is identified with the matter 
lost by massive asymptotic giant branch stars, which evolve during the 
early phases of globular cluster life (D'Antona et al.\ 2002). 

In this paper we examine the implications of the hypothesis that, in many 
globular clusters, stars were born in two separate events: an initial burst 
(first generation), which gives origin to probably all high and intermediate 
mass stars and to a fraction of the cluster stars observed today, and a second, 
prolonged star formation phase (second generation) in which stars form directly 
from the ejecta of the intermediate mass stars of the first generation. In 
particular, we consider in detail the morphology of the horizontal branch in NGC 
2808 and argue that it unveils the early cluster evolution, from the birth of 
the first star generation to the end of the second phase of star formation. 
This framework  provides a feasible interpretation for the still unexplained 
dichotomy of NGC 2808 horizontal branch, 
attributing the lack of stars in the RR Lyr region to the gap in the helium content
between the red clump, whose stars are considered to belong to the first stellar 
generation and have primordial helium, and the blue side of the horizontal branch,
whose {\it minimum} helium content reflects the helium abundance in the smallest 
mass ($\sim 4$\Msun) contributing to the second stellar generation. This scenario
provides constraints on the required Initial Mass Function, in a way that a great 
deal of remnant neutron stars and stellar mass black holes might have been produced. 

\end{abstract}


\section{Introduction}

Chemical inhomogeneities in globular cluster (GC) stars are giving crucial clues 
on the formation and evolution of this most ancient stellar population. Self--
enrichment mechanisms are being examined as ones of the possible -- and most 
plausible -- causes of abundance spread in many GCs. \\
The question of self--enrichment in GCs was raised for the first time in 
connection with the wide spread in metallicity observed in $\omega$Cen 
\citep{freeman-rodgers1975, freeman-norris1981, mallia-pagel1981}. Recently 
there has been a large amount of work on this problem (f.e., Smith et al.\ 2000), 
an aspect of which is the ability of a cluster to retain the matter lost by 
asymptotic giant branch (AGB) stars.  \citet{gnedin2002}  observe that 
$\omega$Cen is not special among GCs in this respect, and that a few dozen other 
clusters should be capable of doing so. In fact, contrary to a substantial 
uniformity of abundance of heavy elements (apart from the quoted $\omega$Cen), 
light elements that are susceptible to abundance changes from proton-capture 
reactions, such as the pp, CN, ON, NeNa, and MgAl cycles, exhibit star-to-star 
abundance variations in many GCs, far in excess of the modest variations seen in 
halo field stars --see, e.g., \citet{smi87}, Kraft (1994)\nocite{kra94}, and 
Sneden (1999,2000)\nocite{sne99,sne00}. In recent years, observations of these 
abundance spreads at the turnoff  and among the subgiant stars 
\citep[e.g.,][]{gratton2001} have shown that these anomalies must be attributed 
indeed to some process of self--enrichment occurring at the first stages of the 
life of the cluster, starting as soon as all the supernovae have already 
exploded (expelling from the clusters their high velocity ejecta) and the 
massive AGB stars begin to evolve. At an epoch starting $\sim 5 10^7$yr from the 
birth of the first stellar generation, the massive AGBs lose mass through low 
velocity winds, so that it can be reasonably speculated that these winds remain 
inside the cluster. \\
The massive AGB envelopes are the ideal place to manifacture elements through 
nuclear reactions in which proton captures are involved, as they are subject to 
hot bottom burning (HBB) \citep[e.g.][]{ventura2001, ventura2002}, although a 
quantitative reproduction of the  observed abundance spreads, e.g., of the 
oxygen vs. sodium anticorrelation \citep{denis2003, ventura2004}, is still far 
from being available. One important feature of this hypothesis is that the 
chemistry of the stars which are formed in the second generation is not {\it 
random}, but reflects the continuous variation of HBB conditions with the 
decrease of the progenitor evolving mass, until the star formation process ends 
\citep[see the discussion in][]{dantona2003}. Thus the ejecta of these stars can 
be the source of the star-to-star abundance variations.\\
The hypothesis that the ejecta are accreted on the already formed stars 
\citep{dgc1983, thoul2002} has been recently falsified by several observations
--see, for a summary of these problems, \citet{gratt-annualrev}. In this paper 
we further explore the hypothesis that such ejecta continuously form second 
generation stars for a time lasting about 200Myr \citep{dantona2002}. \\
We argue that NGC 2808 is one of the clusters able to retain the AGB ejecta, 
even if to a degree lower than $\omega$Cen, which is capable of retaining at 
least part of SNII ejecta (f.e., Smith et al.\ 2000). Large variations in Na 
abundance have been observed among the stars of this cluster 
\citep{carretta2003}, at all luminosity levels along the  red giant (RG) branch. 
So members of this cluster have suffered pollution by Na--enriched matter, as 
observed in M13,  M5,  M4, M92,  M15,  NGC 6752. We do not have information on 
main sequence abundances, as in NGC 6752 \citep{gratton2001} where a primordial 
origin for chemical inhomogeneities appears the only possible explanation, 
nevertheless the hypothesis of self-- enrichment in NGC 2808 is supported by the 
observed Na-- spread. We expect the most massive and concentrated clusters to be 
in the best conditions to retain mass lost by cluster members. NGC 2808 has a 
central density of log ($\rho$) = 4.9 (\Msun/pc$^3$, Pryor \& Meylan 1993), 
among the largest densities found in Galactic GCs; the total visual magnitude is 
\Mv\ = -- 9.26 \citep{harris1996}, which, assuming a ratio $M$/L$_{\rm v}$ = 3, 
gives a total cluster mass well above $10^6$ \Msun. According to Gnedin et al.\ 
(2002, Fig.\ 2), NGC 2808 would be at present quite able of retaining stellar 
winds of about 15 km s$^{-1}$, the characteristic terminal velocity of AGB winds 
(Loup et al.\ 1993). Since according to \citet{gnedin-ostriker1997} the 
destruction time for this cluster is about 6 Hubble times, present conditions 
should be close to those at formation epoch.\\
The hypothesis that GC stars were formed in two different star formation events 
was intended to explain the presence of long blue tails in the HBs 
\citep{dantona2002}  in GCs with large abundance spread \citep{cat1995}. We have 
shown that blue tails can be the result of the evolution of main sequence stars 
with larger than average (cosmological) helium abundance. These latter stars 
have a mass smaller than average at the giant tip at present age and, for 
similar mass loss along the RG branch, will have a smaller mass (and bluer 
location) on the horizontal branch (HB). Helium enrichment is a natural outcome 
of the AGB evolution \citep{ventura2002}, together with other chemical anomalies 
which can help to explain the observed abundance spreads.  We leave aside the 
question of whether the winds from intermediate mass stars (IMS) can collect in 
the cluster central region, and here give origin to a new stellar generation, 
and concentrate on two other important issues:
 \begin{enumerate} 
    \item we have to specify which initial mass function (IMF) is needed to give 
 origin to enough stars in this second formation phase;
     \item we have to explore more in detail which HB morphologies may derive from 
 this process. In particular, the second star formation process may end when the 
 helium content in the last ejecta (which give birth to stars) is still larger 
 than the primordial helium content. This will produce {\it a gap in the helium 
 content}, between the original population of low mass stars, and the last born 
 low mass stars. Does such a gap produce visible consequences in the 
 distribution of stars in the HR diagram? \end{enumerate}
In this paper we attack these two problems, by extending the approach by 
\citet{dantona2002} to a cluster having not only a blue HB with a long blue 
tail, but also a red HB. We will examine the case of the cluster NGC 2808, whose 
HB stellar distribution shows a well populated red clump, an extended blue HB, 
but it contains only two RR Lyr stars. This cluster has been studied intensively 
in recent years. The only way to reproduce the HB distribution is apparently to 
invoke a bimodal (or possibly multimodal) mass distribution 
\citep[e.g.][]{fusipecci-bellazzini1997, catelan1998}. Differences in age or 
metal content among cluster members should be of a size not supported by 
observations. Bedin et al.\ (2000) make an accurate study of the reddening, 
differential reddening and metallicity spread in the cluster, and conclude that 
the $\sigma$ on [Fe/H] is 0.03 on the Carretta \& Gratton scale (1997) and 0.05 
on the Zinn \& West scale (1984). On the other hand, their CM diagram and the 
one by Walker (1999) leave little doubt on the matter. As for an age difference, 
it should be of about 2--4 Gyr as required by the jump in colour (and so in 
mass, about 0.04 \Msun\ at the cluster metallicity) across the empty RR Lyrae 
region: an occurrence which appears rather unlikely. Besides, as noted by 
Rosenberg, Recio- Blanco \& Garcia-Marin (2004), NGC 2808 is coeval with 
clusters of similar metallicity but much shorter HBs such as NGC 362, NGC 1261, 
NGC 1851. Similarly, not very much of definitive can be said on the role of 
rotation on HB distribution. After the many recent investigations on the matter, 
Sweigart (2002) concludes that: ``the impact of mixing and rotation on the HB 
remains to be clearly established". Therefore we think important to explore new 
models for this complex cluster.\\
We here explore the possibility that the dichotomy in the HB is due to the 
differing helium contents of the clump stars and blue HB stars. This assumption 
is certainly parametric and has no pretence of uniqueness, but it has three 
attractive features: i) we know from the RG branch that the cluster stars show a 
dispersion in the abundances linked to the hot CNO cycle; ii) if we accept that 
the massive AGB stars are responsible for this abundance dispersion, all AGB 
models predict that the AGB winds are helium enriched; iii) the epoch of star 
formation from the AGB winds must necessarily end when massive AGBs are still 
evolving, and this will naturally produce a helium content gap.\\
It is important to stress how the large amount of information on the chemical 
composition of the members of GCs has changed profoundly our perception of these 
stellar systems. The impossibility of a simple solution for their formation and 
evolution - one star generation, a common age, one common composition - obliges 
to look for more complex solutions, which have to consider the detailed history 
of star formation, element production and dynamical evolution, and their 
interactions. While we can make reasonable guesses on element production, we 
lack any real knowledge of the dynamical history of GCs. So in this 
investigation we relied on a possible chemical history of NGC 2808, with only 
conjectures on dynamical processes. 

\section{The globular cluster NGC 2808}

\begin{figure}
\includegraphics[width=13cm]{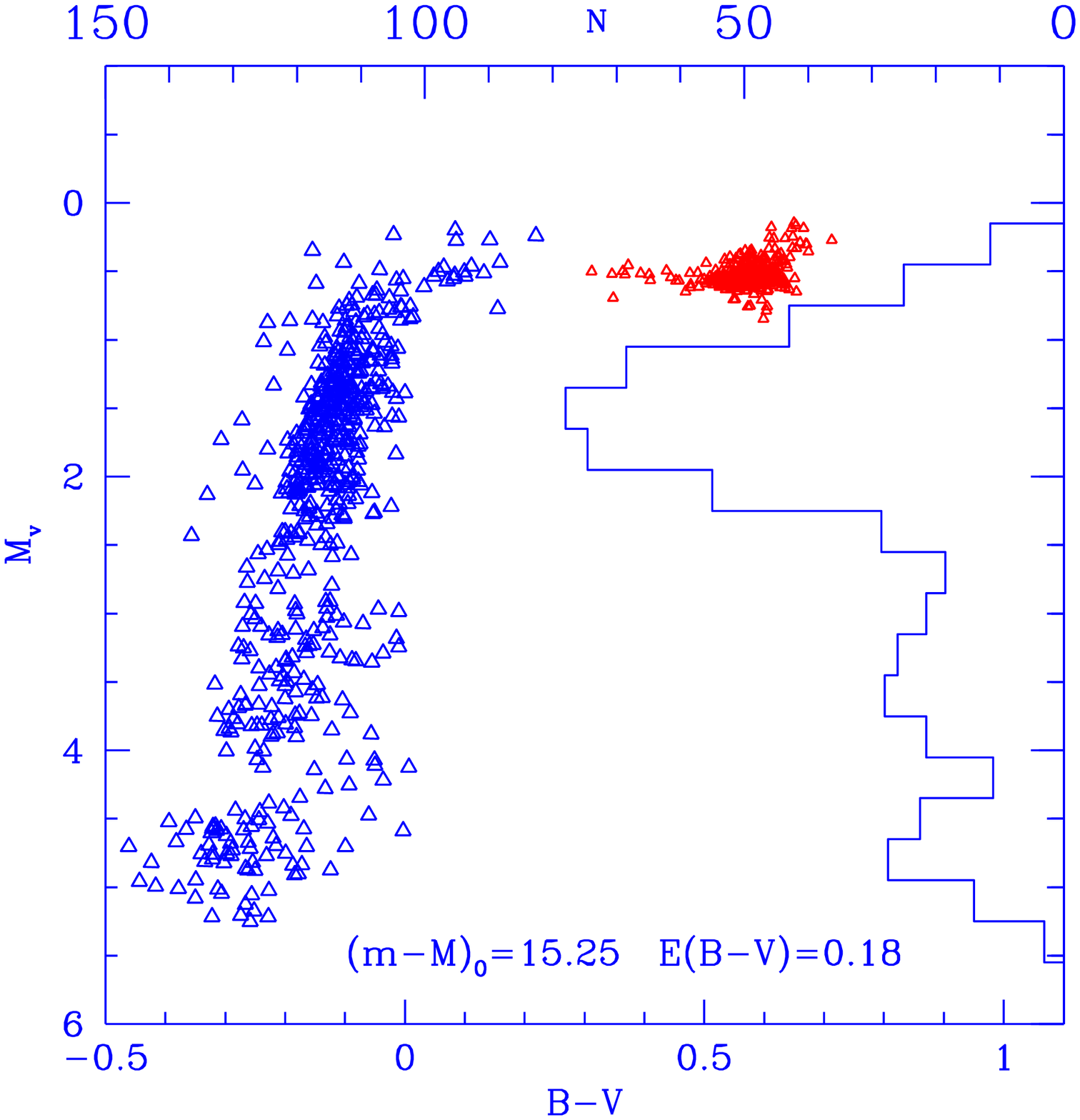}
\figcaption[ngc2808_1.eps]{The HB in the HST data by \citet{bedin2000} for the GC 
NGC 2808. It consists of a red clump plus a composite blue part, along which 
two main gaps are well evident. The histogram (scale given on the top abscissa) 
represents the number of stars in the blue part of the HB, as function of \Mv. 
\label{f1} } \end{figure}

The HB in the CM diagram of NGC 2808 is composed by two regions clearly 
separated: a clump of red stars, similar to those observed in 47 Tuc, M69 and 
M71, and a long branch starting from the blue side of the RR Lyr   gap and 
reaching below the cluster turnoff. In the middle, two RR Lyr variables (one of 
type a,b and one of type c). The (almost) absence of RR Lyr stars is in itself a 
surprise, but other clusters present similar HB distribution, as discussed, for 
example, by \citet{catelan1998}. As mentioned before, they find that the HB 
morphology of NGC 2808 can not be reproduced with a unimodal mass distribution, 
not even  with a large mass dispersion, contrary to the cases of, e.g., NGC 1851 
and NGC 6229. The main difficulty  derives from the large amount of blue HB 
stars: according to \citet{bedin2000}, there are 580 blue HB members and 565 red 
ones\footnote{In this work, we use the numbers given by \citet{bedin2000} for 
the different parts of the HB. These are representative of most of the HB 
population of the cluster. The figures plot the data kindly provided by Bedin, 
but include only the samples whose photometry has been obtained by HST. 
Therefore, e.g., the fit of the red clump stars is based on a sample of 391 
stars.}. The HR diagram from this work is shown in Fig.\ \ref{f1}, where the red 
clump stars and the stars in the three different parts of the blue HB ---EBT1, 
EBT2 and EBT3, according to \citet{bedin2000}. In terms of mass loss along the 
RG branch this fact would imply a strongly bimodal behaviour of the mass loss 
phenomenon, with no apparent reason. \\
The blue HB appears as a sort of  ``different population'', which we can try to 
interpret in terms of self--enrichment and multiple stellar generations in GCs. 
On the basis of the models in Sect 3 for Z=0.001, the first star generation with 
Y = 0.24 and age $\sim$ 13 Gyr reaches the RG tip with a mass of about 0.82 
\Msun\ (in absence of mass loss). The red HB clump appears centered at \0BMV\ 
$\sim$ 0.60 (as shown in  Fig.\ 1, taken from \citet{bedin2000}, assuming E(B--
V)=0.18), which corresponds to a HB mass of 0.68--0.71 \Msun. So the implied 
mass loss is of about 0.13 \Msun. On the other hand, a stellar population with Y 
= 0.28 and age 13 Gyr has at the RG tip a mass of $\sim$ 0.77 \Msun, while for Y 
= 0.32 the RG tip mass is $\sim$ 0.72 \Msun. With the same amount of mass loss, 
a mixture of these two populations would populate the region on the blue side of 
the RR Lyr gap, reaching \Mv\ about 1.5 mag. If we assume as mass loss law 
either Reimers' formulation, or any other formulation currently employed (e.g. 
Lee, Demarque \& Zinn 1994, Catelan 2000), the mass loss rate increases with 
decreasing evolving mass. The giants with Y = 0.28 and Y = 0.32 would lose about 
0.034 \Msun\ and 0.068 \Msun\ more, respectively. This would allow to reach, for 
the case Y = 0.32, \Mv\ $\sim$ 3.6: of course these are order-of-magnitude 
estimates, that show the potentiality of the assumptions, which will be further 
explored with numerical simulations.

\section{The models}

\begin{figure}
   \centering
   \includegraphics[width=13cm]{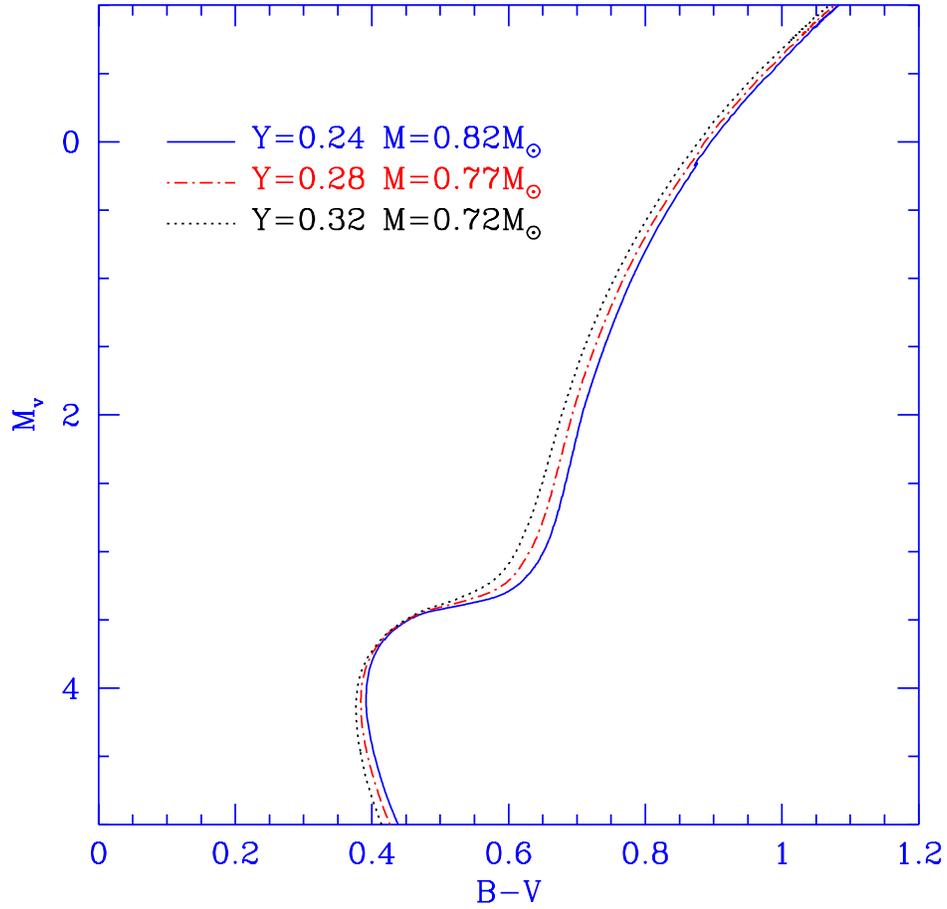}
   \figcaption[tracks.eps]{Tracks of different helium content, corresponding to
   an average age of 13Gyr at the giant branch, are compared in the observational plane.
   The maximum difference in the location occurs at the subgiant branch ($\sim 0.04$mag).
   The turnoff location is not sensibly affected, while the main sequence shows again a color
   difference of few hundreths of magnitude.
    \label{f2}} 
	  \end{figure}

\begin{figure}
   \includegraphics[width=13cm]{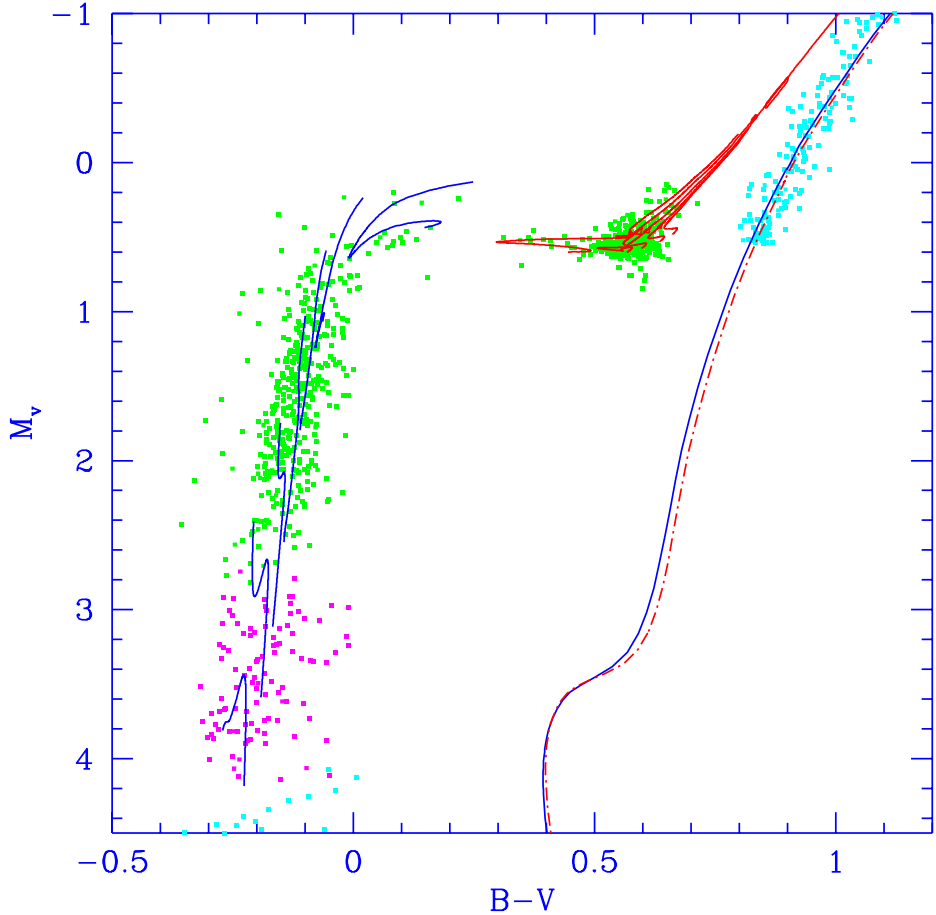}
\figcaption[ngc2808_1.eps]{The models are compared with the observations of the cluster stars 
(distance modulus (m--M)$_0$=15.25 and E(B--V)=0.18).
The turnoff and red giant branch isochrones are shown for 13Gyr and for Y=0.24 and Y=0.28. The 
tracks on the red clump are for masses M=0.66, 0.68, 0.70, 0.74 and 0.8\Msun, Y=0.24.
The tracks on the blue are for M=0.50, 0.52, 0.54, 0.56, 0.58, 0.60 and 0.63\Msun, starting from 
the lowest one. The extreme blue clump (not shown in the figure) is touched by models of 0.48\Msun 
(but see discussion in the text).
\label{f3}} 
\end{figure}

We complemented the models of Z=0.001 computed in paper I for Y=0.24 and Y=0.28 
with models with Y=0.32.  We also evolved red giants with Y=0.32 up to the 
helium flash in order to find the appropriate helium core mass for the HB 
models. In Figure \ref{f2} we show the evolutionary tracks of 0.82\msun\ 
(Y=0.24), 0.77\msun\ (Y=0.28) and 0.72\msun (Y=0.32) which evolve to the helium 
flash in approximately 13Gyr. We can see that the color difference at the 
turnoff is negligible. A color difference of less than 0.04 mag (corresponding 
to $\sim$100K) is found in the subgiant branches, while the RG branches seem to 
converge in color. This simple comparison shows that we do not expect to easily 
recognize a helium spread as large as 0.08 for clusters of this metallicity, 
especially if there is a dominant population with about half of the stars having 
Y=0.24, while the other half are distributed in helium content, as we will show 
in Figure \ref{f6}. Of course a helium variation can be more easily recognized 
in two cases:
     \begin{itemize} 
     \item if in a cluster there are at least two homogeneous consistent fractions of 
stars with different helium abundance, the main sequences and the subgiants may 
show a bifurcation in colors. This might be the case shown by \citet{bedin2004} 
for the main sequence of $\omega$\ Cen, although the peculiarities shown by this 
``cluster" (multiple turnoffs and subgiant branches) depict a much more complex 
situation indeed;
     \item if the dominant population has a high helium content (this could be the 
case, e.g., for the clusters having a fully blue HB, namely those like M13 
showing a clear ``second parameter" effect), it could be recognized by its 
smaller $\delta(B-V)$\ between the turnoff and the RG branch: e.g. a GC having 
normal helium would have $\delta(B-V) \simeq 0.04$mag larger than a GC having 
Y=0.32 and the same age. We mention tentatively that the different morphology of 
the HR diagram of M3 and M13, which has been attributed to the younger age of M3 
\citep{rey2001}, could be due to the fact that M13 stars have a higher helium 
abundance.
     \end{itemize}
We postpone to another paper the discussion of the turnoff and main sequence 
luminosity and mass function in the presence of a helium spread, which requires 
a more detailed analysis, and may also lead to revise the mass function of low 
mass stars in GCs having a helium spread.\\
We further computed models for the evolution up to the RG branch tip and on the 
HB for a total heavy element content of \Z\ = 0.002 and helium abundances \Y\ of 
0.24 and 0.30.  Evolutionary values for the helium core at the giant branch tip 
and for the helium increase in the envelope (first--dredge up) have been 
estimated for an age of 13 $10^9$ yr. \\
The code used is ATON2.0 \citep{ventura1998}, and the transformation from the 
theoretical plane are from \citet{bessell-castelli-plez1998}. The HB models have 
been evolved until the disappearance of helium in the convective core. 
Considering these models together with the models presented in 
\citet{dantona2002}, the mass at the helium tip as function of helium and metal 
content can be approximated by:
\begin{eqnarray}
\log M_{RG}/M_{\odot} \simeq [-0.282+0.092(Y-0.24)]\times \log t   \nonumber \\
    -1.693(Y-0.24)+2.768+12(Z-10^{-3})
\label{eq1} 
\end{eqnarray}
Estimates of the heavy element content Z of NGC 2808 are between 0.001 and 0.002 
(Walker 1999, Bedin et al.\ 2000, Carretta et al.\ 2003).  We compared the HR 
diagram of the RGs and HB of NGC 2808 from \citet{bedin2000} with our sets
of models for Z=0.001 and Z=0.002. The comparison is shown in Figure 
\ref{f3}  for Z=0.001: the red clump, the RG branch and the 
blue HB are well reproduced by the tracks. We adopt E(B--V)=0.18 (Walker 1999, 
Bedin et al.\ 2000) and (m-M)$_0$=15.25.
On the contrary, the same comparison made for Z=0.002 does not reproduce at the 
same time the location of the blue HB and of the RG branch, since the red 
giant tracks are redder by $\sim$0.08mag than the observed branch, once fixed 
the location of the blue HB. This latter depends very weakly on the metallicity and 
on the helium content, and is independent from the treatment of convection. The 
red giant branch and clump locations, on the contrary, depend on the convection 
model. We choose the models with  Z=0.001 on the basis of the better overall
fit of the HR diagram, consistent both with the observed values of metallicity and
with the estimated reddening. We examined also the case Z=0.002; the modeling of the HB
dichotomy follows the same pattern, except that the red clump requires a fine tuning of the
mass loss. In any case, the results for the shape of the IMF turn out to be substantially 
confirmed.\\
For the HB, we computed masses down to $\sim$0.48\Msun, which extend to \Teff 
$\sim 30000$K. These models can not explain the hot subdwarfs which populate the 
most extreme clump of the HB in NGC2808. In fact, together with $\omega$Cen, M54 
(Rosenberg et al.\ 2004), and NGC6388 (Busso, Piotto \& Cassisi 2004),  NGC 2808 
presents the phenomenon of extremely blue, sub--luminous HB members 
\citep{dcruz2000, brown-sweigart2001}. These are objects with a luminosity lower 
than predicted by standard HB models, that is, models with a helium core mass 
determined by RG evolution terminated by the helium flash at the giant tip. 
\citet{castellani2} discovered that HB structures can derive also by a helium 
flash taking place on the white dwarf cooling sequence, and the resulting 
luminosity can be up to 0.1 mag higher than found with standard models 
\citep{dcruz1996}. Still, the sub--luminous stars in $\omega$Cen and NGC 2808 
are up to 0.7 mag fainter than the standard HB, and so not understood in terms 
of HB structures. \citet{sweigart1997a} suggested that a late helium flash along 
the white dwarf cooling sequence can penetrate inside the hydrogen burning 
shell, whose weakness does not provide a sufficiently strong entropy barrier 
\citep{iben1976}. The result would be a structure with an envelope rich in 
helium and carbon, with an effective temperature around 37000 K 
\citep{brown-sweigart2001}, while the hottest HB models lie around 31000 K 
\citep{castellani2, dcruz1996}. The difference in bolometric correction and the 
reduced atmospheric opacity below 912 $\AA$\ produce the observed gap in visual 
magnitude. This means that these objects cannot be modeled with precision. We 
will assume that some of the evolving red giants suffer a delayed helium flash 
followed by mixing, as proposed by \citet{sweigart1997a}, and will attribute 
this sequence of events to all the red giants having the highest Y --and thus 
the smallest initial mass.

\section{The helium distribution in the HB of NGC 2808}

The mass on the HB is:
\begin{equation}
M_{HB} =  M_{RG}(Y,Z) - \Delta M
\label{eq2}
\end{equation}
where $M_{RG}$\ is given by Eq. \ref{eq1}. $\Delta M$\ is the average mass lost 
during the RG phase, with a dispersion $\sigma$ to be determined. Simple 
simulations show that it is possible to interpret the red part of the HB as 
formed by stars with helium content Y=0.24, and the blue part as due to a 
population with enrichment in helium. We can derive important information by 
examining independently the distribution of stars in the red clump and along the 
blue side of the HB.

\subsection{The red clump}
\begin{figure}
\centering
   \includegraphics[width=13cm]{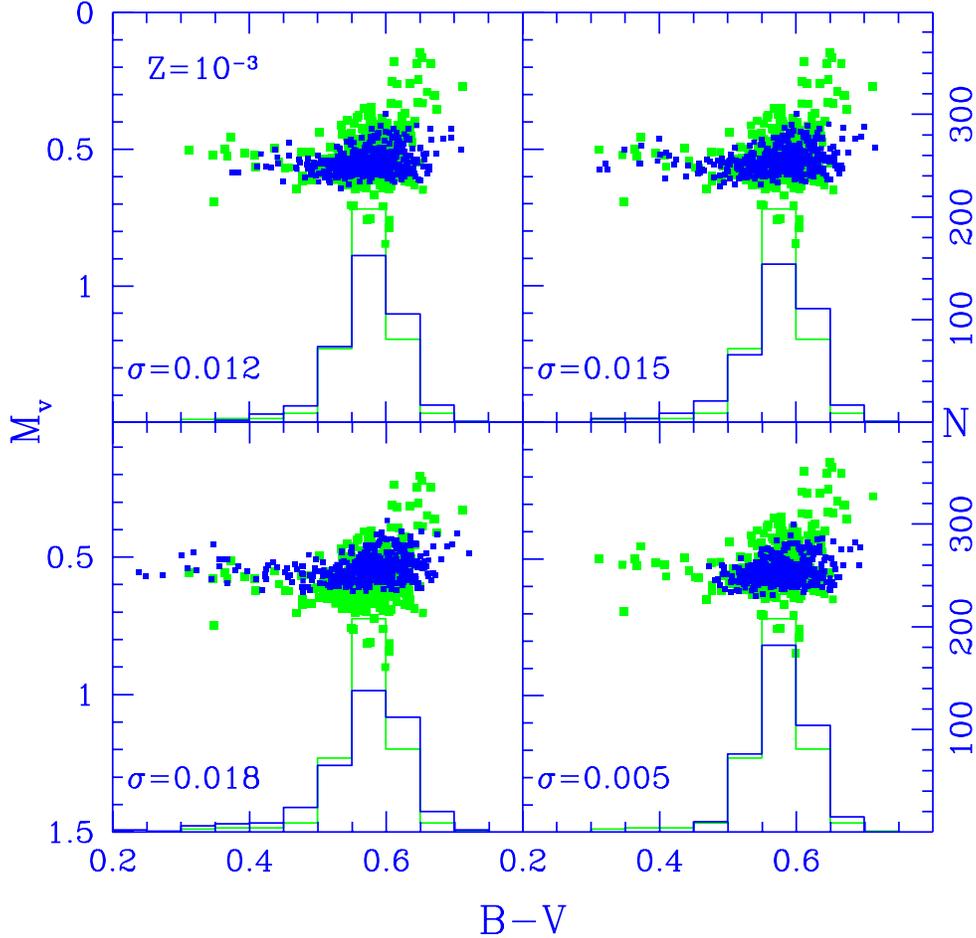}
\figcaption[ngc2808_1.eps]{Simulations for the red clump of the HB of NGC 2808. 
The assumed age is 13Gyr. The gaussian spread $\sigma$\ added to the 
average mass loss of 0.13\Msun along the RGB is indicated in the panels. The 
simulations are compared with the distribution of the clump stars. The distance 
modulus and reddening  are (m$-$M)$_0$=15.25 and E(B--V)=0.18. The 
ordinates on the right give the number of stars in the histograms. Mass spreads from
0.012\Msun\ to 0.018\Msun\ are consistent with the observed lack of stars in the RR Lyrae gap. 
\label{f4}} \end{figure}

Once metallicity and age have been fixed, we adjust the average mass loss along 
the RG branch, and its dispersion, to fit the morphology of the HB red clump. 
The four panels in Fig.\ \ref{f4} show simulations assuming an age of 13 
Gyr. Similar results are obtained for different values of the age, provided that 
we adopt a smaller average $\Delta M$\ to reproduce the red clump, if we adopt a 
larger age. The dispersion assumed around $\Delta M$\ is relatively small: a 
$\sigma \simeq 0.015$\Msun\ is adequate. The red clump is so compact that a 
larger spread in the mass loss tends to populate also the blue side of the 
clump, contrary to the observations. On the basis of  Figure \ref{f4} we choose  
$\sigma=0.015$\Msun.

\subsection{The blue Horizontal Branch} 

\begin{figure}
\includegraphics[width=13cm]{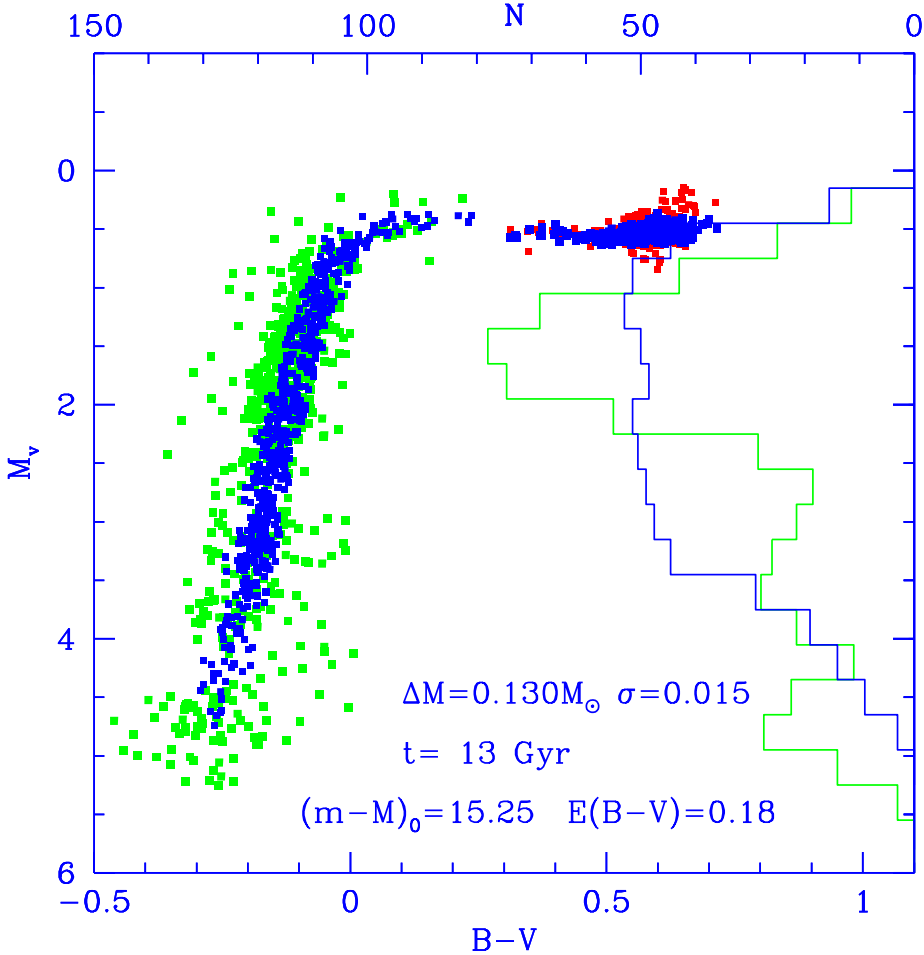}
\figcaption[ngc2808hb1.eps]{
Typical simulation obtained assuming an age of 
13Gyr. The average mass loss on the RGB has been assumed 0.13\Msun\,
in order to reproduce the red clump. 
The corresponding blue HB distribution, assuming 
random helium content between 0.27 and 0.32 is shown. 
Notice that the difference in magnitude 
between the red clump and the red side of the blue HB ( this latter being about 
0.08mag more luminous) is consistent with the increase in the HB luminosity due 
to the helium difference between the two sides. 
 \label{f5}} \end{figure}

Once fitted the red clump, we can suppose that the same $\Delta M$ and the same 
dispersion apply also to the blue side of  the HB, but that the location of the 
stars along this branch depends on the helium content. In other words, {\it the 
mass along the blue HB varies mainly through the variation of $M_{RG}$(Y), as 
given by Eq. \ref{eq1}}. Thus, to obtain the lack of stars in the RR Lyr region 
we need to suppose that the helium content on the reddest side of the blue HB is 
discontinuously larger than the helium content of the red clump.  One full 
simulation for an age of 13 Gyr is shown in Figure \ref{f5}. The clump is 
reproduced as in Figure \ref{f4}, while for the blue side we assume that the 
helium content is randomly distributed between Y=0.27 and Y=0.32.  The minimum 
Y=0.27 is necessary  in order for the RR Lyr region to remain unpopulated. In 
this simulation we attribute maximum observational errors of  $\delta M_{\rm v} 
=0.025$ mag and $\delta$(B--V)=0.035mag. Notice that our purpose here is not to 
obtain a ``perfect" reproduction of the HR diagram, but to obtain clues on the 
helium variation needed to explain the blue HB.\\
An extra bonus of this interpretation of the HB morphology is that it gives an 
explanation of why the reddest stars of the blue HB are slightly more luminous 
than the red clump stars: in fact the different luminosities result naturally 
from the slight difference in the helium content of the red clump and of the red 
side of the blue HB. Incidentally, we remind that there are also some metal rich 
GCs showing a substantial blue HB population, such as NGC 6388 and NGC 6441 
\citep{rich-sosin1997}. One of the debated problems about these clusters is that 
the horizontal part of their HB  shows a marked slope (the bluer stars being 
more luminous). This slope may be attributed to the same mechanism described 
here: the luminosity of the HB increases with \teff\ just because the helium 
abundance of the stars increases.\footnote{The anomalous distribution of the 
period of the RR Lyrae in NGC 6388 and NGC 6441, which does not allow to fit 
them into the Oosterhoff type classification \citep{pritzl2000}, may also have 
the same explanation.}\\
The figures also show the histogram of the number densities obtained in the 
simulations compared to the number densities of the  blue HB, as function of the 
absolute visual magnitude. Two consequences follow: 
\begin{itemize}
    \item we need some additional hypothesis to populate the faintest clump. 
    In particular, we can suppose that some of the stars with the largest helium content 
    (that is, with the smallest initial mass progenitor) actually follow the paths 
   described by \citet{sweigart1997a} and populate the hottest part of the    
blue HB.  
    \item the hypothesis of a random flat distribution in helium does not 
    reproduce well the number densities: we can then reverse the problem, as we 
    show in the next Section. 
\end{itemize}

\subsection{The progenitor masses of the matter forming the blue HB} 
\begin{figure}
\includegraphics[width=13cm]{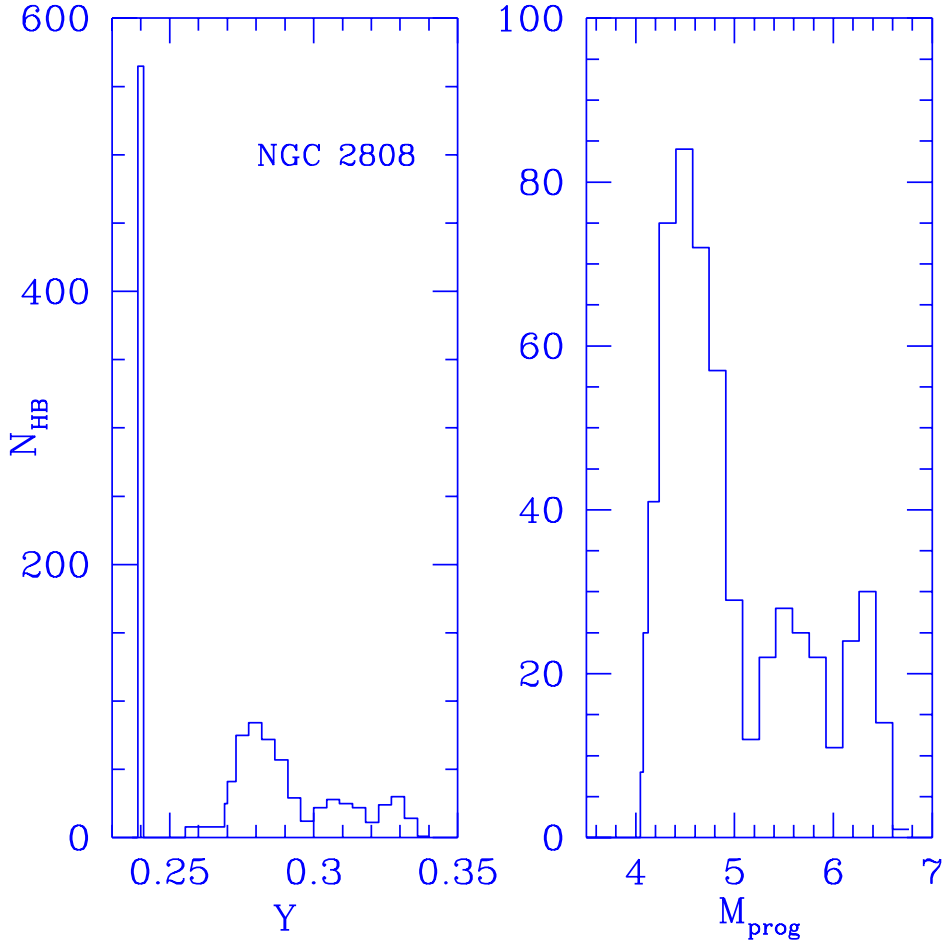}
\figcaption[]{Number of stars versus helium abundance (left) and corresponding 
range of initial mass progenitor (right) in the HB of NGC 2808. The assumptions 
are: 1) that the red clump comes from a population with the initial helium 
abundance Y=0.24; 2) that the location of stars along the blue side of the HB is 
due to the same process of mass loss (average mass loss and spread) which is 
consistent with the red clump and an age of 13Gyr, but the total mass is also a 
function of the initial helium content of the stars, according to equation 
\ref{eq1}; 3) that the material forming the blue HB stars has the helium content of 
the progenitor AGB stars, according to the relation \ref{eq2}. 
\label{f6} } \end{figure}

We can use the relative populations along the HB to infer how many stars 
correspond to a given helium content, for a fixed age and average mass loss, as 
shown in the left side  Fig.\ \ref{f6} for an age of 13 Gyr\footnote{While the 
total number of stars in the red clump is comparable to the number of stars in 
the blue HB, it is evident that the sample having the cosmological helium is 
prominent, in comparison with  the distribution of the objects with larger helium.
For this reason, other evidences for differing helium content among the stars in this
cluster are not easy to be traced: in particular 1) the location of the red giant bump 
\citep[e.g.][]{zoccali1999}, which also depends on the helium abundance, will be 
dominated by the red giants having a normal helium; 2) the parameter R, namely the
ratio between the number of HB stars and the number of RGs above the HB level, is not
much affected, as the level of the HB is taken at the red side, populated by stars
with normal helium, and the evolutionary times of the blue HB stars do not change
in a sensible way with Y.}. We assumed an average 
relation between the helium content and the absolute V magnitude, based on the 
results of the simulation of Figure \ref{f5}. The number of stars - helium 
content distribution on the left side of Figure \ref{f6} can be transformed into 
a relation between the number of stars now on the blue HB and the original AGB 
progenitor star mass, if we know the helium content of the ejecta as function of 
the progenitor mass. From  the results by \citet{ventura2002}, we adopt a simple 
linear relation between the average helium content in the ejecta and the 
progenitor mass\footnote{Actually the helium yields of AGB stars plotted in Fig.\ 
4 of \citet{ventura2002} show a monotonic, but non linear behaviour as function 
of the progenitor mass, and even this characteristic could have a role in 
producing a clumpy distribution of the blue HB in this or in other GCs. 
Nevertheless, it is more instructive to simplify the issue, until we have a 
better understanding of the problem.}. We use:
\begin{equation} 
M_{\rm prog}=3.5 + (Y-0.26)/0.016
\label{eq3}
\end{equation}
The right side of  Fig.\ \ref{f6} shows the result: the blue HB contains stars 
whose progenitors range from $\sim$4  to $\sim$6.5\Msun. Of course this estimate is very 
rough: in particular, we have made the hypothesis that the population of the 
extreme clump can be obtained by extrapolating the average relation between 
helium content and absolute visual magnitude: this produces a ``helium gap" 
which is not real, as we have seen discussing the hottest subdwarf models in Sect. 3.

\section{Deriving a consistent initial mass function}

We should now try to understand what are the implications of Fig.\ \ref{f6} for 
the early evolution of this globular cluster. First it is necessary is to 
understand the mass requirements. The present day mass function of GC stars can 
be studied with sufficient precision only up to the turnoff, which is not larger 
than 0.9\msun. We describe the IMF as made up of power laws having a form:
\begin{equation}
 {dN \over dM} = c \times M^{-(1+x)}
\label{eqx}
\end{equation}
where c is a constant depending on the total mass or number of stars. In the 
whole range up to 0.7\msun, \citet{piotto-zoccali1999} find that the MFs of 
many GCs have very mild slopes (x from $\sim -0.5$\ to $\sim 0.2$). The MF above 
0.7\msun is steeper, and although it is difficult to determine it with care, it 
can be assumed close to Salpeter's slope (x=1.3). Kroupa (2001) makes the 
case for an universal IMF for stellar systems of different metallicities, at 
intermediate and high masses. His reassessment of the IMF is made for many 
stellar populations and different mass ranges--- and shows that the typical IMF 
index is $x \simeq$1.3, that is Salpeter's index, or a larger value, while the 
small masses have much milder slopes. Simplifying Kroupa's results, we describe 
the IMF of GCs as made up of a power law with index x=0 below 0.9\msun, 
connected to a power law with index x=1.3 or x=2 above a ``transition mass" 
\Mtrans=0.9\msun. \\
If   NGC 2808 is made up by a first stellar population having standard 
helium content Y=0.24, from which, today, the stars populating the red clump of 
the HB evolve (and not considering for the moment the loss of stars from the cluster), 
the number of clump stars provide a point of calibration of the IMF. We can 
write:
\begin{equation}
N_{redHB} = \left( {dN \over dM}\right)_{M_{RG}} 
\times \left({dM \over dt} \right)_{M_{RG}} \times \tau_{HB}
\label{eqnhb}
\end{equation}
where $\tau_{HB}$\ is the typical evolutionary time in the HB phase. From 
Eq. \ref{eq1} we can derive the time derivative of the mass $M_{RG}$
\begin{equation}
\left({dM \over dt} \right)_{M_{RG}} = -0.282\times M_{RG} / t_{GC}
\end{equation}
where $t_{GC}$\ is the age of the cluster. As expected, the fraction of stars on 
the HB is proportional to the ratio $\tau_{HB}/t_{GC}$. At the mass today 
evolving on the RGB, given by Eq. \ref{eq1}, and fixing an age of 13Gyr, we can 
calibrate the coefficient in the IMF of Eq. \ref{eqx}. The resulting global 
IMFs, satisfying this constraints and the change of slope at \Mtrans=0.9\msun, 
are shown by the dash--dotted lines in Fig.\ \ref{f7}. \\
Let us assume that indeed all the mass lost by the intermediate mass stars can 
form other stars. Their IMF must have a slope which is at least Salpeter's 
slope, in the hypothesis that  Kroupa's results can be representative also of 
the IMF of stars of mass M$>$\Mtrans\ for the primitive GC. The main feature to 
be explained is EBT1, the most populated clump in the blue HB, which, for an age of 13 
Gyr, and average mass loss during the RG phase of 0.13\Msun, corresponds 
to progenitor masses from \Mprogl=4.1 to \Mprogu=5\Msun. In the sample by 
\citet{bedin2000}  these stars are $N_{BHB}$=401. 
To understand how much mass we need  from these progenitors, 
we have now to make a hypothesis about the distribution of stellar masses in 
this new stellar population coming out from the ejecta. It is difficult to infer 
anything about the formation --if any-- of massive stars, but we can make the 
reasonable hypothesis that, at low masses, this MF must again resemble the MF 
which we see today in GCs: x=0 for $M < M_{trans2}$, x=1.3 for  $M >M_{trans2}$, 
and $ M_{trans2}=0.9$\msun. 
We further make the hypothesis that star 
formation in this case is limited to stars not more massive than 1.5\Msun. From 
this sample, some stars are evolving today in the  main blue clump EBT1. 
The expected number is given by Eq. \ref{eqnhb}, where we now consider the 
number of stars in EBT1, $N_{BHB}=401$. The derivative 
$dM/dt$\ is now slightly different:
\begin{equation}
{dM \over dt} = [-0.282+0.092(Y-0.24)]\times M_{RG} / t_{GC}
\end{equation}
We use Y=0.28, which is the average value for this population (see Fig.\ \ref{f6}).
In this way, we calibrate the coefficients which enter in the IMF of the second 
stellar generation, and establish which is the total mass which must be locked 
into the second generation stars in order to form the appropriate number of blue HB
stars. \\
When we calibrate the IMF of the upper mass range of the first stellar population,
we find that the mass function at masses 4.1--5 must be much larger than predicted
if  the IMF of the GC were ``normal", as described above.   \\
In order to fit both the red clump IMF and the intermediate mass stars IMF, we 
need to use two very different slopes: we adopt Salpeter's slope (or a larger 
slope) for the masses which include our normalization point at 4.1--5\msun, 
and slopes which can  fit the red clump for small masses. This defines a 
\Mtrans,  where the two IMFs must coincide. Some results are shown in Fig.\ 
\ref{f7}, and in Table \ref{tab1}.\\
Finally, it is possible to satisfy both constraints by using a unique IMF. In 
this case it must have a slope x=--0.14, as we see in Table \ref{tab1}. Such an 
IMF can not be extrapolated too much, as it predicts an unrealistic high mass 
for the cluster and is not consistent with all the evidence concerning the IMFs 
derived from many different stellar environments. 

\subsection{The extreme blue tails 2 and 3}
Although the number of stars in the EBT1 provides the best constraint to our 
IMF, we can try to use the number of stars in the EBT2 and EBT3 together, to 
infer another point of the star formation rate of the first stellar generation. 
Of course the relation between progenitor mass and helium content becomes more 
uncertain here, as we do not know well the evolutionary meaning of the EBT3 
stars. We can consider two cases: either these stars come from progenitor masses 
between 5.1 and 6.6\msun\ (as indicated in Fig \ref{f6}), or we assign to all of 
them the mass range 5.1--6\msun\, that is, we attribute all stars to the mass 
range of the EBT2, and consider the EBT3 clump as due to random occurrence of 
the flash--induced mixing suggested by \citet{sweigart1997a}. The corresponding 
mass function points are given in Fig \ref{f7}. We see that in both cases the 
mass function is well below the predictions based on the point relative to EBT1, 
and assuming IMF slopes up to x=2. We do not give too much weight to this 
result, but at face value it indicates that the second generation star formation 
is less favoured at higher AGB masses. This may be due, e.g., to the last 
supernova events which take place into the cluster, when already the most 
massive AGB stars are evolving.

\begin{deluxetable}{rrrrrrrr} 
\tablecaption{IMF and Mass of the primordial Globular Cluster NGC
2808} \tablecolumns{8}
\tablehead{
\colhead{M$_{tot}^{(a)}$} & \colhead{M$_{trans}$}& \colhead{x$_{low}$}&
\colhead{x$_{up}$}& \colhead{N$_{tot}^{(a})$}& \colhead{N$_{NS}$}&
\colhead{N$_{BH}$}& \colhead{M$_{initial}^{(b)}$} }
\startdata
 1.01E+06  &  0.90 &  0.00 &  1.30 &  1.04E+06 &  1.40E+04 &  3.90E+03 &  1.01E+06 \\
 5.58E+05  &  0.90 &  0.00 &  2.00 &  9.52E+05 &  2.62E+03 &  3.16E+02 &  5.58E+05 \\
 5.37E+07  &  5.00 & -0.14 & -0.14 &  2.78E+06 &  4.99E+05 &  8.82E+05 &  5.37E+07 \\
\tableline
 7.77E+06  &  3.03 & -0.50 &  1.01 &  1.65E+06 &  1.86E+05 &  7.39E+04 &  1.04E+07 \\
 8.07E+06  &  2.20 & -0.99 &  1.01 &  1.68E+06 &  1.86E+05 &  7.39E+04 &  1.04E+07 \\
 8.26E+06  &  1.80 & -1.50 &  1.01 &  1.73E+06 &  1.86E+05 &  7.39E+04 &  1.04E+07 \\
 8.38E+06  &  1.58 & -2.00 &  1.01 &  1.78E+06 &  1.86E+05 &  7.39E+04 &  1.04E+07 \\
 8.47E+06  &  1.44 & -2.50 &  1.01 &  1.82E+06 &  1.86E+05 &  7.39E+04 &  1.04E+07 \\
\tableline
 5.86E+06  &  3.23 & -0.50 &  1.30 &  1.58E+06 &  1.47E+05 &  4.08E+04 &  1.01E+07 \\
 6.26E+06  &  2.41 & -0.99 &  1.30 &  1.64E+06 &  1.47E+05 &  4.08E+04 &  1.01E+07 \\
 6.54E+06  &  1.98 & -1.50 &  1.30 &  1.74E+06 &  1.47E+05 &  4.08E+04 &  1.01E+07 \\
 6.74E+06  &  1.74 & -2.00 &  1.30 &  1.83E+06 &  1.47E+05 &  4.08E+04 &  1.01E+07 \\
 6.89E+06  &  1.57 & -2.50 &  1.30 &  1.92E+06 &  1.47E+05 &  4.08E+04 &  1.01E+07 \\
\tableline
 3.91E+06  &  3.55 & -0.50 &  2.00 &  1.49E+06 &  8.48E+04 &  1.02E+04 &  1.63E+07 \\
 4.50E+06  &  2.79 & -0.99 &  2.00 &  1.62E+06 &  8.48E+04 &  1.02E+04 &  1.63E+07 \\
 5.02E+06  &  2.34 & -1.50 &  2.00 &  1.81E+06 &  8.48E+04 &  1.02E+04 &  1.63E+07 \\
 5.46E+06  &  2.05 & -2.00 &  2.00 &  2.03E+06 &  8.48E+04 &  1.02E+04 &  1.63E+07 \\
 5.83E+06  &  1.85 & -2.50 &  2.00 &  2.24E+06 &  8.48E+04 &  1.02E+04 &  1.63E+07 \\
\tableline
 3.10E+06  &  3.80 & -0.50 &  3.00 &  1.43E+06 &  4.13E+04 &  1.59E+03 &  5.46E+07 \\
 3.88E+06  &  3.15 & -0.99 &  3.00 &  1.62E+06 &  4.13E+04 &  1.59E+03 &  5.46E+07 \\
 4.73E+06  &  2.71 & -1.50 &  3.00 &  1.94E+06 &  4.13E+04 &  1.59E+03 &  5.46E+07 \\
 5.56E+06  &  2.40 & -2.00 &  3.00 &  2.31E+06 &  4.13E+04 &  1.59E+03 &  5.46E+07 \\
 6.37E+06  &  2.18 & -2.50 &  3.00 &  2.73E+06 &  4.13E+04 &  1.59E+03 &  5.46E+07 \\
\enddata 
\noindent
\tablenotetext{a}{Mass and total number of stars included between 0.08\Msun
and 100\Msun, using the IMF indices of column 3 and 4}
\tablenotetext{b}{Mass which the cluster had, if it were constrained by
the total wind mass between 4.1 and 5\Msun, but the IMF extends till
M=0.9\Msun with index $x_{up}$, changing to $x_{low}=0$\ below M=0.9\Msun.}
\label{tab1}
\end{deluxetable}

\section{Discussion}

\begin{figure}
\includegraphics[width=13cm]{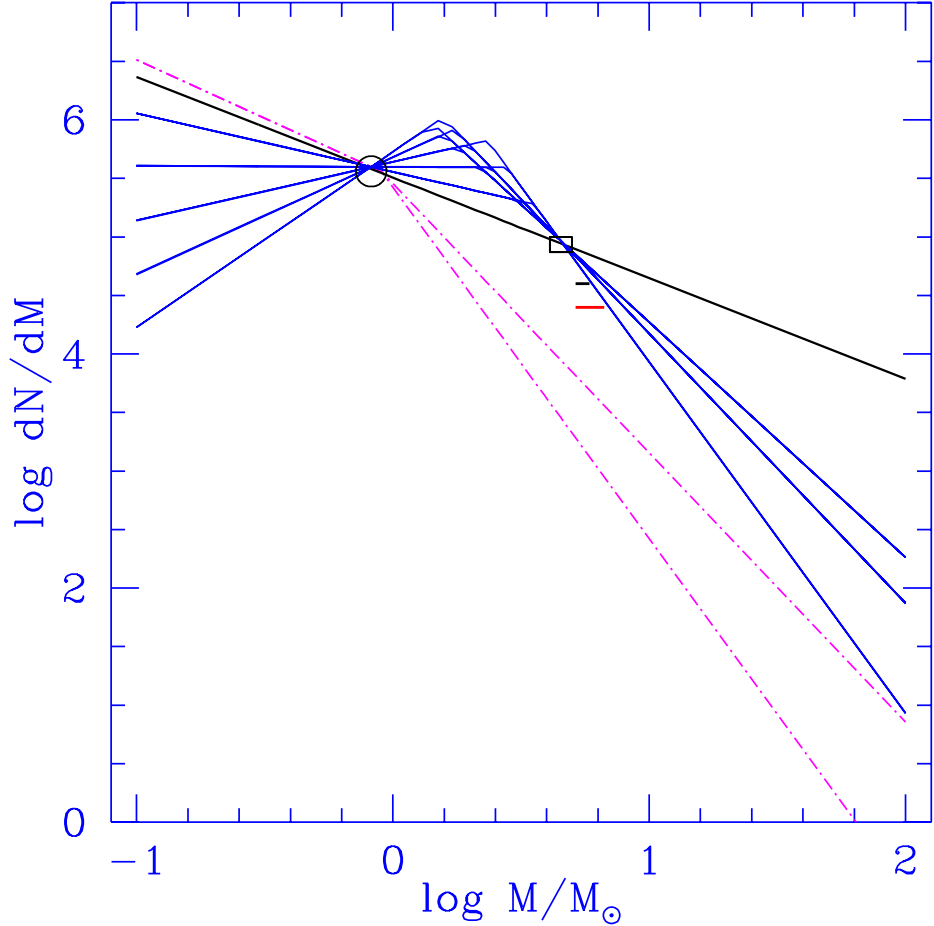}
\figcaption[imfs]{Possible choices for the IMFs of the first generation stars in 
NGC 2808, based on the constraints of the HB morphology. The circle at M$\simeq 
0.82$\Msun\ refers to the number of the red clump stars, $N_{clump}=565$\ from 
\citet{bedin2000}. The square at M$\simeq 4 - 5$\Msun\ refers to the number of 
stars in the EBT1 ($N_{BHB}=401$). The continuous 
straight line passing through the circle and the rectangle is a unique mass 
function of slope x=$-$0.14 (see Table \ref{tab1}). 
The two dash--dotted lines are IMFs built with x=0 at 
M$<$0.9\Msun, and x=1.3 or x=2 at M$>$0.9\Msun, following the experimental 
determination of the IMF in the Galaxy \citep{kroupa2001}. 
At M=5.1-6\msun\ and M=5.1-6.6\msun\ we show the IMF points based on the
number of stars in EBT2 and EBT3 taken together (see text).\label{f7} }
\end{figure}

Figure \ref{f7} summarizes the schematic IMFs  based on our interpretation of 
the morphology of the HB in NGC 2808. The result has been obtained by many 
simplifying assumptions, but a variation of the input hypotheses does not 
produce a qualitatively different behaviour, as we have also checked by assuming 
a different global metallicity for the cluster.

1)  Let us assume that the original IMF of NGC 2808 is similar to the 
``universal" IMF by \citet{kroupa2001}, in a simplified form: two power laws 
with indices x=0 below 0.9\msun\ and x=1.3 above it.  In this case all the HB 
stars must come from only one population, as the matter lost by IMS is short by 
a factor 10 or so from being able to form the population which we see today in 
the blue HB. A similar conclusion was reached a long time ago by 
\citet{cottrell-dacosta} and was in fact the most severe problem in any kind of 
self--enrichment scheme for GCs ---apart from the self--pollution by accretion 
on stars proposed by \citet{dgc1983}. 

2) In order to have enough stars in the main blue clump EBT1,  we need a {\it 
relatively flat} IMF, that is  x$\sim 0$.   But is there any evidence for a flat 
IMF up to the mass range of massive AGBs, that is stars of 4-7\msun ? A possible 
solution is that the IMF slope in the central parts of clusters is indeed 
different from that in the field. \citet{panagia2000}, in a systematic study of 
the young stellar population around the Supernova 1987A, show that the number 
ratio of low mass (1-2\msun) stars to massive stars is very low, $\sim$1.4, 
within 130$\farcs$ from the supernova, corresponding to about 30pc, while it is 
$\sim$8.9 in the general field: this translates in an IMF slope x=0.5, to be 
compared with x=1.5 for the entire field. An IMF of mild slope is therefore not 
to be ruled out: it may well be that in compact stellar systems the IMF slope is 
mild, while it is much steeper in the general field, reflecting different 
processes of star formation, as \citet{panagia2000} suggest. Other hints came 
some years ago based on the possible interpretation of MACHO halo objects in 
terms of halo white dwarfs \citep{adams1996, chabrier1996}. The same 
\citet{kroupa2001} suggests a systematically varying IMF, which, in the past, 
had a characteristic fragmentation mass around 1--2\msun, while today it is 
about 0.1\msun.

3) Actually,  the process of formation of a second generation of stars can not 
be decoupled from the dynamical evolution of the cluster:  during the first 
phases of life, the IMS had time to segregate into the cluster central parts, 
where, for this reason, indeed the IMF would look like those shown in Fig.\ 
\ref{f7}\footnote{Indeed, for a young GC belonging to the Large Magellanic 
Cloud, NGC 1866, which has an age close to the age for which we expect the 
second stage of star formation to have been efficient in GCs, 
\citet{brocato2003} find a strong evidence for mass segragation in the cluster 
core, for masses $\sim 4$\Msun.}. The ejecta of these stars gave birth to the 
second generation, but the external parts of the cluster, containing all the 
other  first generation low mass stars, was lost, leaving the present proportion 
of first and second generation stars. The  IMFs shown in Fig.\ \ref{f7} can be 
considered then simple examples of what actually should be obtained by 
considering the proper dynamic evolution of the cluster, including the loss of 
low mass stars from it.   What seems to be clear is that, if the true IMF of the 
first stellar population of GCs was similar to the field IMF, we {\it need that 
most of the low mass stars, today progenitors of the HB red clump, has been lost 
from the cluster}.

\subsection{The neutron stars and black holes from the first generation} 

Table \ref{tab1} lists some properties derived from the IMFs shown in Fig.\ 
\ref{f7}, which refer to an age of 13Gyr, an average mass loss of 0.13\msun\ and 
a $\sigma$\ of 0.015\msun. In particular, we see that the initial mass of NGC 
2808 must have been much larger indeed than the mass we see today. Of course, 
this work does not enter into the problems of loss of stability of the cluster, 
which however will be a powerful constraint on the reliability of each proposed 
IMF.  We wish to point out an additional possible consequence of our 
interpretation of the HB of NGC 2808: it provides an appealing hint for the 
solution of a problem in GC stellar content: the amount of neutron stars (NS) 
and black holes (BH).\\
We can formally compute the number of NSs and BHs for each of the IMFs which 
satisfy the constraint posed by the blue HB, in the hypothesis that we are 
authorized to extrapolate their number to the mass range of NS and BH 
progenitors. The total number of NS is obtained by considering the progenitors 
between the upper mass which gives origin to AGB stars (7\Msun) and 25\Msun, and 
BH formation is considered between 25 and 100\Msun. \\
The large number of millisecond pulsars (MSPs) discovered in recent years in 
many GCs, and in particular the 20 MSPs discovered in the cluster 47 Tuc 
\citep{camilo2000} indicate that the richest GCs may contain more than $\sim 
1000$\ NSs \citep[e.g.,][]{pfahl2003} a number embarassingly large. In 
fact, the number of NSs predicted by a normal IMF in NGC 2808 would be between 
2600 and 14000 (top of Table \ref{tab1}), implying a retention factor from 7 
to 40\%, at variance with the characteristic kick speeds of single radiopulsar 
in the Galaxy, which have an average $<v> \sim 250 - 300$Km/s  
\citep{hp1997},  much larger than the escape velocity from typical clusters
($\sim 25$Km/s). 
\citet{pfahl2003} have shown that the retention factor is between 1\% and 8\%, 
even including consideration of binariety, but it would have been much smaller 
if they had taken into account a number of other effects. Table \ref{tab1} shows 
that the retention problem becomes less severe if we assume that the IMF of 
intermediate mass stars, necessary to interpret the blue HB morphology, can be 
extended to the masses which form neutron stars, even if the slope is much 
larger than Salpeter's (up to x=2).\\
The presence of stellar or intermediate mass BHs in Globular Clusters is 
exciting today a renewed interest. On the one side, recent observations by STIS 
on HST of the velocity dispersion in the center of the cluster M15 seem to 
indicate the possible presence of a massive BH (a few thousand \msun) in the core 
\citep{gerssen2002}. On the other hand, the presence of an intermediate mass BH, 
or of binary BH systems, may explain the peculiar dynamical behaviour of the 
MSPs in the cluster NGC 6752 \citep{damico2002}, and in particular the location 
of PSR A at more than 3 half--mass radii from the center \citep{colpi2003}. \\
According to Table \ref{tab1}, an important number of stellar mass black holes 
could have been formed in the cluster, with interesting consequences on its 
further dynamical evolution. BHs should have minor retention problems, as the 
galactic distribution of black hole X-ray binaries has a mean velocity of only 
$\sim$40 km s$^{-1}$ (White \& van Paradijs 1996), implying no or negligible 
kicks. However, close (three-- or four--body) dynamical encounters among BHs, 
may expell them from clusters, and in fact current N-body simulations (Portegies 
Zwart \& McMillan 2000) foresee no black hole or at most one black hole binary 
to remain in the cluster. In addition, even less is known about their detailed 
formation \citep[see, e.g.][]{fryer-kalogera2001}, and we have inferred their 
possible numbers by extrapolating the IMF to masses an order of magnitude larger 
than our calibration point at $\sim 5$\msun. The numbers in Table \ref{tab1}, 
therefore, should be taken as helpful indication, and certainly not as 
predictions. \\
As a final caveat, notice that the interaction between the AGB winds and the 
stellar mass BHs possibly present in the central regions is not devoid of 
consequences, either as a help to the star formation mechanism, or as a sink of 
gas: the problem of the early evolution of GCs is all to be worked out.

\section{Conclusions} 

In this paper we have used the galactic GC NGC 2808 as a laboratory to 
experiment on recent ideas on the early evolution of GCs. The HB in this cluster 
shows a very peculiar morphology: it is well populated in the red clump and in 
the blue and extremely blue region, but almost devoid of stars in the RR Lyr 
region. This morphology can be understood in terms of the coexistence of two 
stellar generations, separated by a difference between the helium content of the 
first one and the minimum helium content of the second one ($\Delta Y \simeq 
0.03$).  The small luminosity difference between the red and blue side of the 
horizontal part of the HB is also explained by this helium difference.
\\
The peculiar HB morphology of NGC 2808 is therefore consistent with the 
hypothesis that the stars in many GCs were formed in two different events, a 
hypothesis primarily invoked to explain the abundance spreads in the light 
elements involved in the hot CNO cycle \citep{cottrell-dacosta, dantona2002}. 
Using an average relation between the progenitor AGB mass and the helium content 
of its ejecta and the HB data by \citet{bedin2000}, we derive two points of the 
IMF of the cluster stars: 1) the point at M=0.82\msun, that is the mass function 
value for the stars which today are found in the red clump,  having the 
primordial helium content; 2) the point for the mass range $\sim 4.1 - 5$\msun , 
whose ejecta should populate the most populous clump of the blue HB. Although 
this second point is somewhat model dependent, it clearly results that we need 
many more AGB stars than predicted by an IMF similar to those inferred for many 
stellar environments \citep{kroupa2001}. We favour then the possibility that 
many of the low mass stars of the first stellar generation have been lost, but a 
dynamical study of these first stages of star formation is needed.
\\
We comment on the large number of NSs --and possibly also BHs-- which our model 
predict, and which would make easier to understand the population of millisecond 
pulsars in many GCs and the possible presence of intermediate mass BHs in the 
clusters M15 and NGC 6752.
\\
As for possible checks of the proposed scenario, the disappearance of surface 
helium (through sedimentation) at temperatures at which it would become 
observable, makes direct measurements impossible. A way to check the multiple 
star generations hypothesis could be the accurate measurement of main sequence 
masses in double stars, as well as the detailed analysis of the behaviour of 
various CM diagram features, such as the magnitude difference between the RG 
branch bump and the turnoff.

\begin{acknowledgements}
We thank all the GC italian friends who helped us in collecting data and 
ideas, and mainly R. Gratton, G. Piotto, M. Zoccali and L.R. Bedin. `Rolly' Bedin 
is also thanked for a critical reading of the manuscript. F.D.  thanks A. 
Possenti and M. Colpi for useful discussion about the neutron star and black 
hole content of galactic Globular Clusters. Comments by an anonymous referee
helped us to clarify the proposed scenario.
\end{acknowledgements}

\begin{thebibliography}{}

\bibitem[Adams \& Laughlin (1996)]{adams1996} Adams, F..~C.~\& 
Laughlin, G.\ 1996, \apj, 468, 586 

\bibitem[Bedin et al. (2000)]{bedin2000}
Bedin, L.~R., Piotto, G., Zoccali, M., Stetson, P.~B., Saviane, I., Cassisi,
S., \& Bono, G.\ 2000, \aap, 363, 159

\bibitem[Bedin et al. (2004)]{bedin2004}
Bedin, L.~R., Piotto, G., Anderson, J., Cassisi, S., King, I. R., Momany, Y., Carraro, G.\ 2004,
\apj, 605, L125

\bibitem[Bessell, Castelli, \& Plez(1998)]{bessell-castelli-plez1998} 
Bessell, M.~S., Castelli, F., \& Plez, B.\ 1998, \aap, 333, 231 

\bibitem[Brocato et al. (2003)]{brocato2003} Brocato, E., 
Castellani, V., Di Carlo, E., Raimondo, G., \& Walker, A.~R.\ 2003, \aj, 
125, 3111 

\bibitem[Brown et al. (2001)]{brown-sweigart2001} Brown,
T.~M., Sweigart, A.~V., Lanz, T., Landsman, W.~B., \& Hubeny, I.\ 2001, \apj,
562, 368

\bibitem[Busso, Piotto, \& Cassisi(2004)]{2004MmSAI..75...46B} Busso, G., 
Piotto, G., \& Cassisi, S.\ 2004, Memorie della Societa Astronomica 
Italiana, 75, 46 


\bibitem[Camilo et al. (2000)]{camilo2000}
Camilo, F., Lorimer, D. R., Freire, P., Lyne, A. G., \& Manchester, R. N.
2000, \apj, 535, 975

\bibitem[Carretta \& Gratton (1997)]{carretta1997} Carretta E., \& Gratton R.G. 1997, A\&AS, 121, 95 

\bibitem[Carretta et al. (2003)]{carretta2003} Carretta, E., Bragaglia, A., Cacciari, 
C., \& Rossetti, E.\ 2003, \aap, 410, 143

\bibitem[Castellani \& Castellani (1993)]{castellani2} 
Castellani. M, \& Castellani V. 1993, ApJ, 407, 649

\bibitem[Catelan \& de Freitas Pacheco (1995)]{cat1995} 
Catelan, M., \& de Freitas Pacheco, J. A. 1995, A\&A, 297, 345

\bibitem[Catelan et al. (1998)]{catelan1998} Catelan,
M., Borissova, J., Sweigart, A.~V., \& Spassova, N.\ 1998, \apj, 494, 265

\bibitem[Catelan(2000)]{2000ApJ...531..826C} Catelan, M.\ 2000, \apj, 531, 
826 

\bibitem[Chabrier et al. (1996)]{chabrier1996} 
Chabrier, G., Segretain, L., \& M'era, D.\ 1996, \apjl, 468, L21 

\bibitem[Colpi, Mapelli, \& Possenti(2003)]{colpi2003} Colpi, M., 
Mapelli, M., \& Possenti, A.\ 2003, \apj, 599, 1260 

\bibitem[Cottrell \& Da Costa (1981)]{cottrell-dacosta}
Cottrell, P.~L.~\& Da Costa, G.~S.\ 1981, \apjl, 245, L79

\bibitem[D'Amico et al. (2002)]{damico2002} D'Amico, N., Possenti, 
A., Fici, L., Manchester, R.~N., Lyne, A.~G., Camilo, F., \& Sarkissian, 
J.\ 2002, \apjl, 570, L89 

\bibitem[D'Antona, Gratton, \& Chieffi (1983)]
{dgc1983} D'Antona, F., Gratton, R., \& Chieffi, A.\ 1983, Memorie della
Societa Astronomica Italiana, 54, 173

\bibitem[D'Antona et al. (2002)]{dantona2002} D'Antona,
F., Caloi, V., Montalb{\' a}n, J., Ventura, P., \& Gratton, R.\ 2002, \aap,
395, 69

\bibitem[D'Antona (2003)]{dantona2003} D'Antona, F.\ 2003, Societa 
Astronomica Italiana Memorie Supplement, 3, 64 

\bibitem[D'Cruz et al. (1996)]{dcruz1996} 
D'Cruz N.L., Dorman B., Rood R.T., \& O'Connell R.W. 1996, ApJ, 466, 359

\bibitem[D'Cruz et al. (2000)]{dcruz2000} D'Cruz, N.~L.~et al.\ 
2000, \apj, 530, 352 

\bibitem[Denissenkov \& Herwig (2003)]{denis2003}
Denissenkov, P.~A.~\& Herwig, F.\ 2003, \apjl, 590, L99

\bibitem[Freeman \& Rodgers (1975)]{freeman-rodgers1975}
Freeman, K. C.; Rodgers, A. W.\ 1975, \apj, 201, L71

\bibitem[Freeman \& Norris (1981)]{freeman-norris1981}
Freeman, K. C.; Norris, J.\ 1981, \araa, 19, 319

\bibitem[Fryer \& Kalogera (2001)]{fryer-kalogera2001}
Fryer, C. L. \& Kalogera, V. 2001, \apj, 554, 548

\bibitem[Fusi Pecci \& Bellazzini(1997)]{fusipecci-bellazzini1997} Fusi Pecci, 
F.~\& Bellazzini, M.\ 1997, The Third Conference on Faint Blue Stars, 255 

\bibitem[Gerssen et al. (2002)]{gerssen2002}
Gerssen, J., van der Marel, R. P., Gebhardt, K., Guhathakurta, P.,
Peterson, R. C., \& Pryor, C. 2002, AJ, 124, 3270 (addendum 125, 376
[2003])

\bibitem[Gnedin \& Ostriker(1997)]{gnedin-ostriker1997} 
Gnedin, O.~Y.~\& Ostriker, J.~P.\ 1997, \apj, 474, 223 

\bibitem[Gnedin et al. (2002)]{gnedin2002} Gnedin, O.~Y., Zhao, H., 
Pringle, J.~E., Fall, S.~M., Livio, M., \& Meylan, G.\ 2002, \apjl, 568, 
L23

\bibitem[Gratton et al.(2001)]{gratton2001} Gratton,
R.~G.~et al.\ 2001, \aap, 369, 87

\bibitem[Gratton, Sneden \& Carretta (2004)]{gratt-annualrev}
Gratton, G., Sneden, C., \& Carretta, E. 2004, \araa, in press

\bibitem[Hansen \& Phinney(1997)]{hp1997} Hansen, B.~M.~S.~\& 
Phinney, E.~S.\ 1997, \mnras, 291, 569 

\bibitem[Harris (1996)]{harris1996} Harris, W.~E.\ 1996, VizieR 
Online Data Catalog, 7195

\bibitem[Iben (1976)]{iben1976}
Iben, I.Jr. 1976, \apj, 208,165

\bibitem[Kraft (1994)]{kra94}
Kraft, R. P. 1994, PASP, 106, 553

\bibitem[Kroupa (2001)]{kroupa2001} Kroupa, P.
2001, MNRAS, 322, 231

\bibitem[Lee, Demarque, \& Zinn(1994)]{lee1994} Lee, Y., 
Demarque, P., \& Zinn, R.\ 1994, \apj, 423, 248 

\bibitem[Loup, Forveille, Omont, \& Paul (1993)]{loup1993} Loup, 
C., Forveille, T., Omont, A., \& Paul, J.~F.\ 1993, \aaps, 99, 291 

\bibitem[Mallia \& Pagel (1981)]{mallia-pagel1981} Mallia, E.~A.~\& 
Pagel, B.~E.~J.\ 1981, \mnras, 194, 421 

\bibitem[Panagia et al. (2000)]{panagia2000} 
Panagia, N., Romaniello, Scuderi \& Kirshner 2000, ApJ 539, 197

\bibitem[Pfahl et al. (2003)]{pfahl2003}
Pfahl, E., Rappaport, S. \& Podsiadlowski, P.\ 2003, \apj, 573, 283

\bibitem[Piotto \& Zoccali (1999)]{piotto-zoccali1999}
Piotto, G. \& Zoccali, M. 1999,  A\&A, 345, 485

\bibitem[Portegies Zwart \& McMillan(2000)]{2000ApJ...528L..17P} Portegies 
Zwart, S.~F.~\& McMillan, S.~L.~W.\ 2000, \apjl, 528, L17 

\bibitem[Pryor \& Meylan (1993)]{pryor-meylan1993}
Pryor, C. \& Meylan, G. 1993, in Structure and Dynamics of Globular Clusters,
S.G. Djorgovsky and G. Meylan eds., ASP Conference Series, 50, 357

\bibitem[Pritzl et al. (2000)]{pritzl2000} 
Pritzl, B., Smith, H.~A., Catelan, M., \& Sweigart, A.~V.\ 2000, \apjl, 530, L41

\bibitem[Rey et al.(2001)]{rey2001} Rey, S., Yoon, S., Lee, Y., 
Chaboyer, B., \& Sarajedini, A.\ 2001, \aj, 122, 3219 

\bibitem[Rich et al. (1997)]{rich-sosin1997} Rich, R.~M.~et al.\ 1997, \apjl, 484, L25

\bibitem[Rosenberg, Recio-Blanco, \& 
Garc{\'{\i}}a-Mar{\'{\i}}n(2004)]{2004ApJ...603..135R} Rosenberg, A., 
Recio-Blanco, A., \& Garc{\'{\i}}a-Mar{\'{\i}}n, M.\ 2004, \apj, 603, 135 

\bibitem[Smith (1987)]{smi87}Smith G. H. 1987, PASP, 99, 67

\bibitem[Smith et al.(2000)]{smith2000} Smith, V.~V., Suntzeff, 
N.~B., Cunha, K., Gallino, R., Busso, M., Lambert, D.~L., \& Straniero, O.\ 
2000, \aj, 119, 1239 

\bibitem[Sneden (1999)]{sne99} Sneden C. 1999, Ap\&SS, 265, 145

\bibitem[Sneden (2000)]{sne00}
Sneden C. 2000, in 35th Liege International Astrophysics Colloquium: 
The Galactic Halo, from Globular Clusters to Field Stars, ed. A. Noels, 
P. Magain, D. Caro, E. Jehin, G. Parmentier, and A. Thoul (Li\`ege Belgium: 
Institut d'Astrophysique et de G\'eophysique), p. 159

\bibitem[Sweigart (1997)]{sweigart1997a}
Sweigart, A.~V.\ 1997, 
Third Conference on Faint Blue Stars, ed. A. G. D. Philip, J. Liebert,
R. Saffer and D.S. Hayes, (Schenectady, L. Davis Press), p.3

\bibitem[Sweigart (2002)]{sweigart2002} Sweigart,  A. V. 2002, 
Highlights of Astronomy, Vol. 12, p. 292

\bibitem[Thoul et al.(2002)]{thoul2002} Thoul, A., Jorissen, A., 
Goriely, S., Jehin, E., Magain, P., Noels, A., \& Parmentier, G.\ 2002, 
\aap, 383, 491 

\bibitem[Ventura et al. (1998)]{ventura1998} 
Ventura, P., Zeppieri, A., Mazzitelli, I., \& D'Antona, F.\ 1998, \aap, 334, 953

\bibitem[Ventura et al. (2001)]{ventura2001} Ventura,
P., D'Antona, F., Mazzitelli, I., \& Gratton, R.\ 2001, \apjl, 550, L65 

\bibitem[Ventura, D'Antona, \& Mazzitelli (2002)]
{ventura2002} Ventura, P., D'Antona, F., \& Mazzitelli, I.\ 2002, \aap, 393,
215

\bibitem[Ventura et al. (2004)]{ventura2004}
Ventura, P., Mazzitelli, I. \& D'Antona, F.\ 2004, Proceedings of the Joint Discussion 04
of the XXIV IAU General Assembly, Sydney, Astrophysical Impact of Abundances in Globular
Cluster Stars, eds. F.D'Antona and G. Da Costa, Mem.S.A.It. 75, n. 2, in presse

\bibitem[Walker (1999)]{walker1999} Walker, A.~R.\ 1999, \aj, 118, 
432 

\bibitem[White \& van Paradijs(1996)]{white-vanparadijs} White, N.~E.~\& 
van Paradijs, J.\ 1996, \apjl, 473, L25 

\bibitem[Zinn \& West (1984)]{zinn84} Zinn R., \& West M.J. 1984, ApJS, 55, 45

\bibitem[Zoccali et al.(1999)]{zoccali1999} Zoccali, 
M., Cassisi, 
S., Piotto, G., Bono, G., \& Salaris, M.\ 1999, \apjl, 518, L49 

\end{thebibliography}
\end{document}